\documentclass[10pt,a4paper]{article}

\usepackage[top=1.3in, bottom=1.3in, left=1.5in, right=1.5in]{geometry}
\usepackage[breaklinks]{hyperref}
\usepackage[T1]{fontenc} 
\usepackage[latin1]{inputenc}
\usepackage{aeguill}
\usepackage{tikz}
\usepackage{amsfonts}
\usepackage{amsmath}
\usepackage{amssymb}
\usepackage{gnuplot-lua-tikz}
\usepackage{subfigure}
\usepackage{url}
\usepackage{garamond}
\usepackage[urw-garamond]{mathdesign}
\usepackage[sort&compress]{natbib}

\newenvironment{keywords}{
       \list{}{\advance\topsep by0.35cm\relax\small
       \leftmargin=1cm
       \labelwidth=0.35cm
       \listparindent=0.35cm
       \itemindent\listparindent
       \rightmargin\leftmargin}\item[\hskip\labelsep
                                     \bfseries Keywords:]}
     {\endlist}

\begin{document}

\title{Observation of large-scale multi-agent based simulations}

\author{
Gildas Morvan$^{1,2}$~~~Alexandre Veremme$^{1,3}$~~~Daniel Dupont$^{1,3}$\\~\\
{\footnotesize $^{1}$Univ Lille Nord de France, F-59000 Lille, France}\\
{\footnotesize $^{2}$UArtois, LGI2A, F-62400, Béthune, France}\\
{\footnotesize $^{3}$HEI, F-59046, Lille, France}\\~\\
{\footnotesize \url{http://www.lgi2a.univ-artois.fr/~morvan/}}\\{\footnotesize \url{gildas.morvan@univ-artois.fr}}\\~\\~\\\includegraphics[width=3cm]{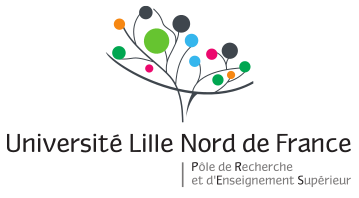}~~\includegraphics[width=3cm]{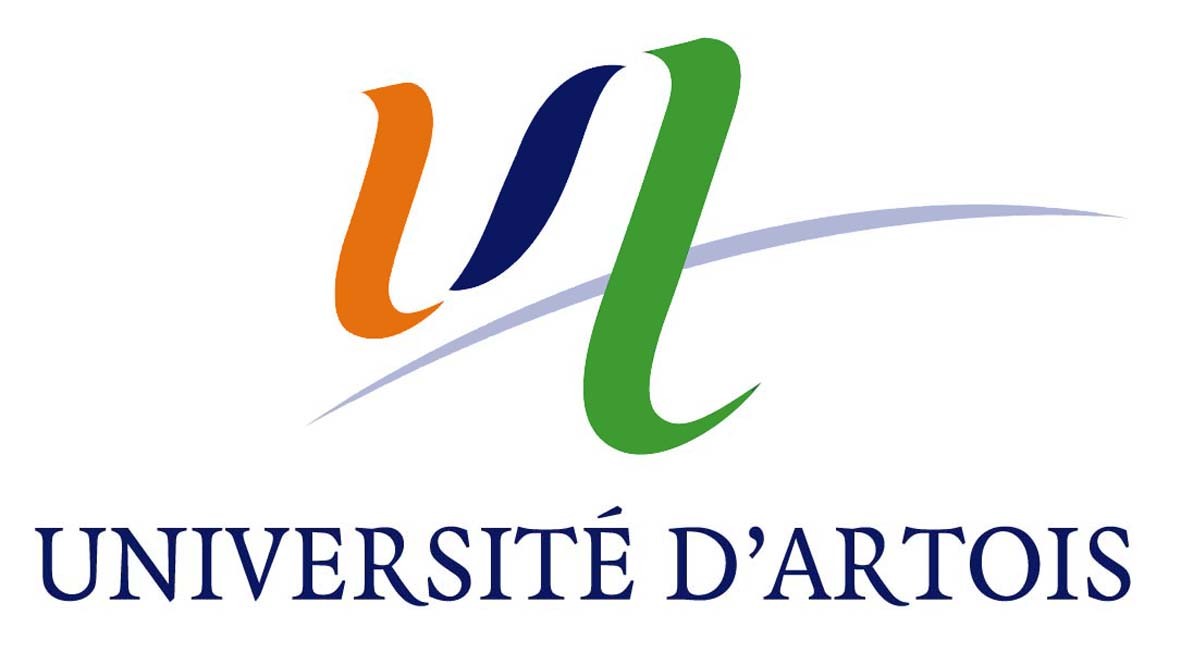}~~\includegraphics[width=2.5cm]{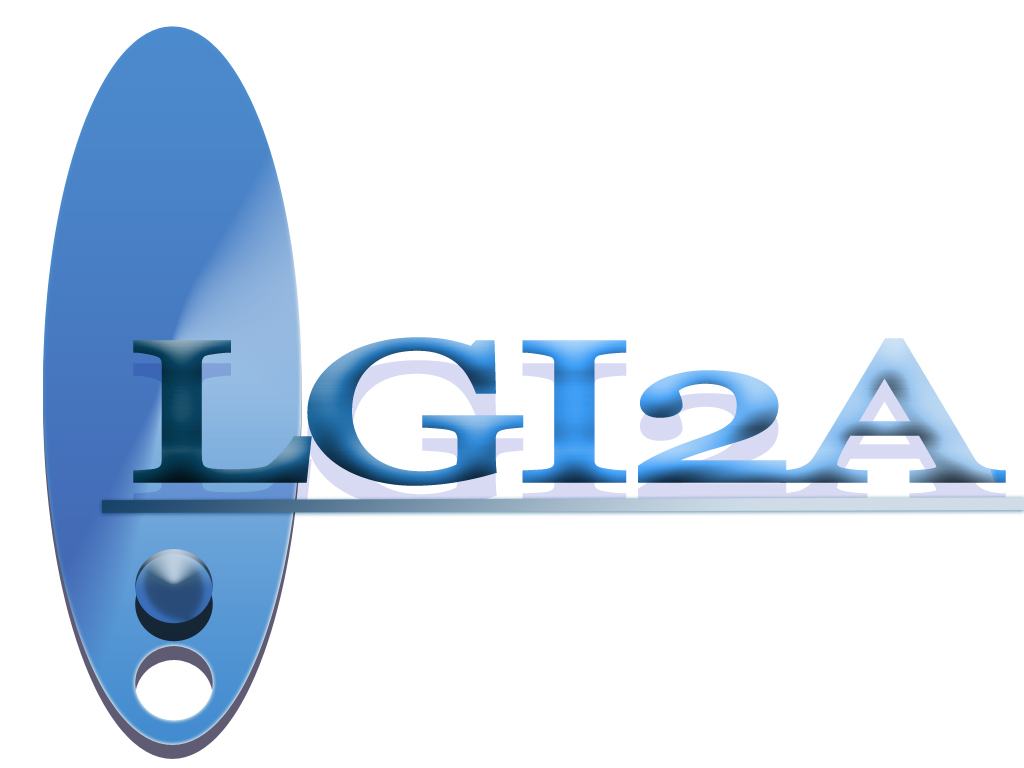}~~\includegraphics[width=4.2cm]{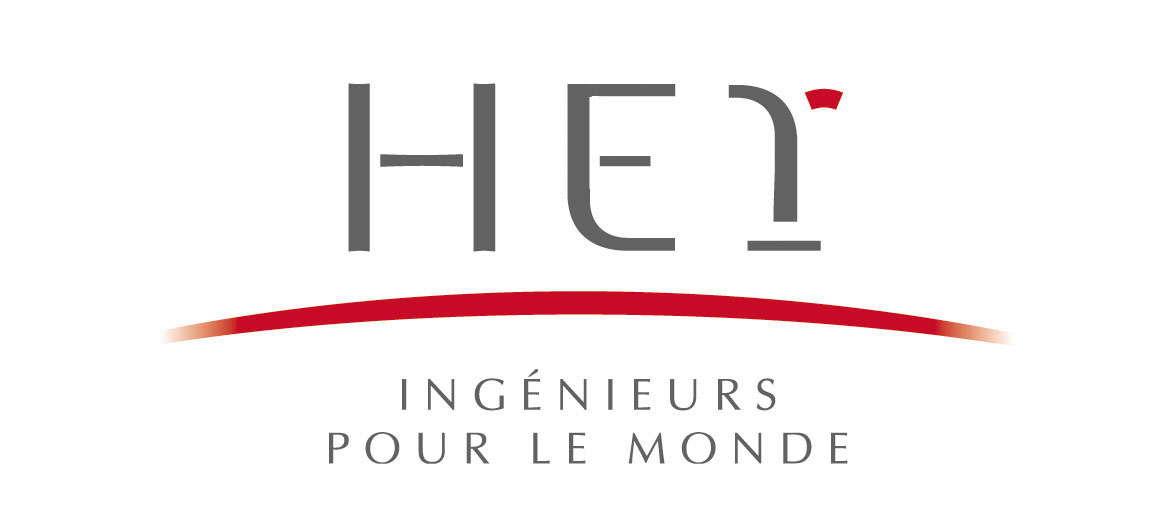}\\~\\
}

\date{}

\garamond

\maketitle

\begin{abstract}

The computational cost of large-scale multi-agent based simulations (MABS) can be extremely important, especially if simulations have to be monitored for validation purposes. In this paper, two methods, based on self-observation and statistical survey theory, are introduced in order to optimize the computation of observations in MABS. An empirical comparison of the computational cost of these methods is performed on a toy problem. 
\end{abstract}

\begin{keywords}
large-scale multi-agent based simulations, observation methods, scalability 
\end{keywords}

\section{Introduction}
\label{}

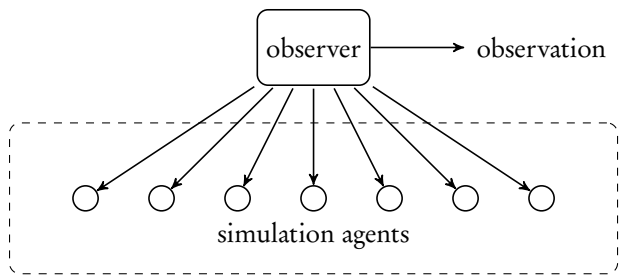
\begin{figure}[ht]
	\begin{center}
\tikzstyle{blocvide}= [minimum width=0.2cm, minimum height=0.2cm,text centered]
\tikzstyle{blocdebase}= [draw, text centered,minimum height=1cm, rounded corners]

\tikzstyle{simulation}= [draw, dashed, text centered,minimum height=2cm, minimum width=8cm, rounded corners]

\tikzstyle{obs} = [blocdebase]

\pgfdeclarelayer{background}
\pgfdeclarelayer{foreground}
\pgfsetlayers{background,main,foreground}

\begin{tikzpicture}[->,>=stealth',shorten >=1pt,auto,semithick]
    
    \node[obs] (observateur) at (4cm, 0cm)  {observer};
    
    \node[blocvide] (observation) at (7cm, 0cm)  {observation};
    
    \path (observateur) edge (observation) {};
    
    \node[blocvide] at (4cm, -2.5cm) {simulation agents};
    
    \foreach \x in {1, 2, 3, 4, 5, 6, 7} {
    	\node [circle,draw]  (a) at (\x cm, -2cm){};
	\path (a) edge [<-]  node {} (observateur);
    }
		
	\begin{pgfonlayer}{background}
		\node[simulation] at (4cm, -2cm){};
   	 \end{pgfonlayer}

\end{tikzpicture}
		\caption{Brute-force direct observation method}
		\label{bruteforce-scheme}
	\end{center}
\end{figure}

Theoretical and practical advances in the field of multi-agent based simulations~(MABS) allow modelers to simulate very complex systems to solve real world problems. However the analysis and validation of simulations remain engineering problems that do not have "turnkey" solutions. Thus, MABS users conducting such tasks face two main issues: 
\begin{enumerate}
\item define validation metrics for the simulation,
\item compute efficiently the metrics. 
\end{enumerate}
The first issue is generally solved by constructing a set of \textit{ad-hoc} qualitative or quantitative rules on simulation properties. To evaluate these rules, it is then mandatory to observe the corresponding simulation properties and thus to consider the second issue. A distinctive characteristic of MABS is that  global simulation properties are not necessary directly observable: they may need to be computed from local agent properties. Fortunately,  most of modern MABS platforms come with observation frameworks and toolboxes. Basically, three types of observation methods are generally available~\citep{Railsback:2006}:
\begin{enumerate}
\item interactive observation: users select observed properties during simulations, \textit{e.g.}, using a point and click interface, 
\item brute-force direct observation: simulation agents sharing a given property are monitored; agent properties are then aggregated by a so-called \textit{observer} agent that computes the observation~(fig.~\ref{bruteforce-scheme}), 
\item indirect observation: the observed property is inferred from the observable consequences of agent actions, \textit{e.g.}, in the environment.
\end{enumerate}

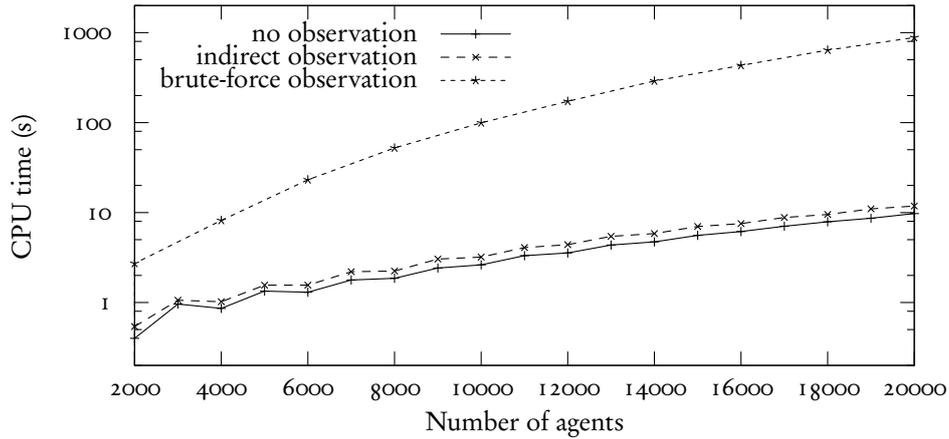
\begin{figure}[ht]
	\begin{center}
		\begin{tikzpicture}[gnuplot]
\gpmonochromelines
\gpcolor{gp lt color border}
\gpsetlinetype{gp lt border}
\gpsetlinewidth{1.00}
\draw[gp path] (1.688,0.985)--(1.778,0.985);
\draw[gp path] (11.947,0.985)--(11.857,0.985);
\draw[gp path] (1.688,1.460)--(1.778,1.460);
\draw[gp path] (11.947,1.460)--(11.857,1.460);
\draw[gp path] (1.688,1.703)--(1.778,1.703);
\draw[gp path] (11.947,1.703)--(11.857,1.703);
\draw[gp path] (1.688,1.819)--(1.868,1.819);
\draw[gp path] (11.947,1.819)--(11.767,1.819);
\node[gp node right] at (1.504,1.819) { 1};
\draw[gp path] (1.688,2.178)--(1.778,2.178);
\draw[gp path] (11.947,2.178)--(11.857,2.178);
\draw[gp path] (1.688,2.652)--(1.778,2.652);
\draw[gp path] (11.947,2.652)--(11.857,2.652);
\draw[gp path] (1.688,2.896)--(1.778,2.896);
\draw[gp path] (11.947,2.896)--(11.857,2.896);
\draw[gp path] (1.688,3.011)--(1.868,3.011);
\draw[gp path] (11.947,3.011)--(11.767,3.011);
\node[gp node right] at (1.504,3.011) { 10};
\draw[gp path] (1.688,3.371)--(1.778,3.371);
\draw[gp path] (11.947,3.371)--(11.857,3.371);
\draw[gp path] (1.688,3.845)--(1.778,3.845);
\draw[gp path] (11.947,3.845)--(11.857,3.845);
\draw[gp path] (1.688,4.089)--(1.778,4.089);
\draw[gp path] (11.947,4.089)--(11.857,4.089);
\draw[gp path] (1.688,4.204)--(1.868,4.204);
\draw[gp path] (11.947,4.204)--(11.767,4.204);
\node[gp node right] at (1.504,4.204) { 100};
\draw[gp path] (1.688,4.563)--(1.778,4.563);
\draw[gp path] (11.947,4.563)--(11.857,4.563);
\draw[gp path] (1.688,5.038)--(1.778,5.038);
\draw[gp path] (11.947,5.038)--(11.857,5.038);
\draw[gp path] (1.688,5.281)--(1.778,5.281);
\draw[gp path] (11.947,5.281)--(11.857,5.281);
\draw[gp path] (1.688,5.397)--(1.868,5.397);
\draw[gp path] (11.947,5.397)--(11.767,5.397);
\node[gp node right] at (1.504,5.397) { 1000};
\draw[gp path] (1.688,5.756)--(1.778,5.756);
\draw[gp path] (11.947,5.756)--(11.857,5.756);
\draw[gp path] (1.688,0.985)--(1.688,1.165);
\draw[gp path] (1.688,5.756)--(1.688,5.576);
\node[gp node center] at (1.688,0.677) { 2000};
\draw[gp path] (2.828,0.985)--(2.828,1.165);
\draw[gp path] (2.828,5.756)--(2.828,5.576);
\node[gp node center] at (2.828,0.677) { 4000};
\draw[gp path] (3.968,0.985)--(3.968,1.165);
\draw[gp path] (3.968,5.756)--(3.968,5.576);
\node[gp node center] at (3.968,0.677) { 6000};
\draw[gp path] (5.108,0.985)--(5.108,1.165);
\draw[gp path] (5.108,5.756)--(5.108,5.576);
\node[gp node center] at (5.108,0.677) { 8000};
\draw[gp path] (6.248,0.985)--(6.248,1.165);
\draw[gp path] (6.248,5.756)--(6.248,5.576);
\node[gp node center] at (6.248,0.677) { 10000};
\draw[gp path] (7.387,0.985)--(7.387,1.165);
\draw[gp path] (7.387,5.756)--(7.387,5.576);
\node[gp node center] at (7.387,0.677) { 12000};
\draw[gp path] (8.527,0.985)--(8.527,1.165);
\draw[gp path] (8.527,5.756)--(8.527,5.576);
\node[gp node center] at (8.527,0.677) { 14000};
\draw[gp path] (9.667,0.985)--(9.667,1.165);
\draw[gp path] (9.667,5.756)--(9.667,5.576);
\node[gp node center] at (9.667,0.677) { 16000};
\draw[gp path] (10.807,0.985)--(10.807,1.165);
\draw[gp path] (10.807,5.756)--(10.807,5.576);
\node[gp node center] at (10.807,0.677) { 18000};
\draw[gp path] (11.947,0.985)--(11.947,1.165);
\draw[gp path] (11.947,5.756)--(11.947,5.576);
\node[gp node center] at (11.947,0.677) { 20000};
\draw[gp path] (1.688,5.756)--(1.688,0.985)--(11.947,0.985)--(11.947,5.756)--cycle;
\node[gp node center,rotate=-270] at (0.246,3.370) {CPU time (s)};
\node[gp node center] at (6.817,0.215) {Number of agents};
\node[gp node right] at (5.534,5.379) {no observation};
\gpcolor{gp lt color 0}
\gpsetlinetype{gp lt plot 0}
\draw[gp path] (5.718,5.379)--(6.634,5.379);
\draw[gp path] (1.688,1.344)--(2.258,1.798)--(2.828,1.741)--(3.398,1.970)--(3.968,1.955)%
  --(4.538,2.117)--(5.108,2.140)--(5.678,2.276)--(6.248,2.318)--(6.818,2.440)--(7.387,2.476)%
  --(7.957,2.581)--(8.527,2.623)--(9.097,2.709)--(9.667,2.760)--(10.237,2.830)--(10.807,2.889)%
  --(11.377,2.936)--(11.947,2.998);
\gpsetpointsize{4.00}
\gppoint{gp mark 1}{(1.688,1.344)}
\gppoint{gp mark 1}{(2.258,1.798)}
\gppoint{gp mark 1}{(2.828,1.741)}
\gppoint{gp mark 1}{(3.398,1.970)}
\gppoint{gp mark 1}{(3.968,1.955)}
\gppoint{gp mark 1}{(4.538,2.117)}
\gppoint{gp mark 1}{(5.108,2.140)}
\gppoint{gp mark 1}{(5.678,2.276)}
\gppoint{gp mark 1}{(6.248,2.318)}
\gppoint{gp mark 1}{(6.818,2.440)}
\gppoint{gp mark 1}{(7.387,2.476)}
\gppoint{gp mark 1}{(7.957,2.581)}
\gppoint{gp mark 1}{(8.527,2.623)}
\gppoint{gp mark 1}{(9.097,2.709)}
\gppoint{gp mark 1}{(9.667,2.760)}
\gppoint{gp mark 1}{(10.237,2.830)}
\gppoint{gp mark 1}{(10.807,2.889)}
\gppoint{gp mark 1}{(11.377,2.936)}
\gppoint{gp mark 1}{(11.947,2.998)}
\gppoint{gp mark 1}{(6.176,5.379)}
\gpcolor{gp lt color border}
\node[gp node right] at (5.534,5.071) {indirect observation};
\gpcolor{gp lt color 1}
\gpsetlinetype{gp lt plot 1}
\draw[gp path] (5.718,5.071)--(6.634,5.071);
\draw[gp path] (1.688,1.500)--(2.258,1.849)--(2.828,1.829)--(3.398,2.049)--(3.968,2.049)%
  --(4.538,2.227)--(5.108,2.236)--(5.678,2.395)--(6.248,2.421)--(6.818,2.547)--(7.387,2.586)%
  --(7.957,2.696)--(8.527,2.733)--(9.097,2.827)--(9.667,2.865)--(10.237,2.946)--(10.807,2.986)%
  --(11.377,3.060)--(11.947,3.098);
\gppoint{gp mark 2}{(1.688,1.500)}
\gppoint{gp mark 2}{(2.258,1.849)}
\gppoint{gp mark 2}{(2.828,1.829)}
\gppoint{gp mark 2}{(3.398,2.049)}
\gppoint{gp mark 2}{(3.968,2.049)}
\gppoint{gp mark 2}{(4.538,2.227)}
\gppoint{gp mark 2}{(5.108,2.236)}
\gppoint{gp mark 2}{(5.678,2.395)}
\gppoint{gp mark 2}{(6.248,2.421)}
\gppoint{gp mark 2}{(6.818,2.547)}
\gppoint{gp mark 2}{(7.387,2.586)}
\gppoint{gp mark 2}{(7.957,2.696)}
\gppoint{gp mark 2}{(8.527,2.733)}
\gppoint{gp mark 2}{(9.097,2.827)}
\gppoint{gp mark 2}{(9.667,2.865)}
\gppoint{gp mark 2}{(10.237,2.946)}
\gppoint{gp mark 2}{(10.807,2.986)}
\gppoint{gp mark 2}{(11.377,3.060)}
\gppoint{gp mark 2}{(11.947,3.098)}
\gppoint{gp mark 2}{(6.176,5.071)}
\gpcolor{gp lt color border}
\node[gp node right] at (5.534,4.763) {brute-force observation};
\gpcolor{gp lt color 2}
\gpsetlinetype{gp lt plot 2}
\draw[gp path] (5.718,4.763)--(6.634,4.763);
\draw[gp path] (1.688,0.985)--(1.688,2.337)--(2.828,2.907)--(3.968,3.446)--(5.108,3.868)%
  --(6.248,4.201)--(7.387,4.486)--(8.527,4.756)--(9.667,4.962)--(10.807,5.165)--(11.947,5.332);
\gppoint{gp mark 3}{(1.688,2.337)}
\gppoint{gp mark 3}{(2.828,2.907)}
\gppoint{gp mark 3}{(3.968,3.446)}
\gppoint{gp mark 3}{(5.108,3.868)}
\gppoint{gp mark 3}{(6.248,4.201)}
\gppoint{gp mark 3}{(7.387,4.486)}
\gppoint{gp mark 3}{(8.527,4.756)}
\gppoint{gp mark 3}{(9.667,4.962)}
\gppoint{gp mark 3}{(10.807,5.165)}
\gppoint{gp mark 3}{(11.947,5.332)}
\gppoint{gp mark 3}{(6.176,4.763)}
\gpcolor{gp lt color border}
\gpsetlinetype{gp lt border}
\draw[gp path] (1.688,5.756)--(1.688,0.985)--(11.947,0.985)--(11.947,5.756)--cycle;
\gpdefrectangularnode{gp plot 1}{\pgfpoint{1.688cm}{0.985cm}}{\pgfpoint{11.947cm}{5.756cm}}
\end{tikzpicture}
		\caption{CPU times (log scale) needed to compute observed and unobserved simulations for $E(Z)=N/5$}
		\label{fig1}
		\end{center}
\end{figure}

While the first method is clearly unadapted to the observation of large-scale or batch simulations, the second has  an important computational cost. This issue is illustrated with a simple case study inspired by "StupidModel"~\citep{Railsback:2005}: $N$ agents move randomly in a two dimensional environment $\mathcal{E}$ discretized into $100 \cdot 100$ square cells with Moore neighbourhood during 1000 steps. An area $\mathcal{Z} \subseteq \mathcal{E}$ is defined. The number of agents $Z$ in the area $\mathcal{Z}$ is observed at each simulation step. This simulation is implemented on the MadKit/TurtleKit platform\footnote{All the simulations and observation methods described in this paper have also been implemented on the MASON platform~\citep{Luke:2005}, leading to similar results.}~\citep{Michel:2005}. Figure~\ref{fig1} shows the CPU times needed to compute unobserved and observed simulations on a Dell Precision 650 workstation\footnote{CPU: 2 $\times$ 3.06 GHz Intel Xeon\texttrademark{}, RAM: 4 $\times$ 1 GB. Full specification:\\ \url{http://www.dell.com/downloads/emea/products/precn/precn_650_uk.pdf}.\\ All the results presented in this paper have been computed by this machine.},  using indirect and brute-force direct observation methods, for the given expected value $E(Z)=N/5$, as a function of the number of simulation agents $N$.

This work is based on existing implementations (by MadKit/TurtleKit and MASON) of the direct observation method. Thus, an empirical computational complexity metric, \textit{i.e.}, the CPU time needed to compute simulations, is used. This metric, denoted $\mathcal{C}$, depends, in our case study, on the number of simulation agents, $N$, and on the expected value of the cardinal of the subset  of simulation agents computed by $filter$, $E(Z)$. 

These results show that, in this case, indirect observation has a minor impact on the computational cost of the simulations.  Kaminka \textit{et al.} also note that this method is not intrusive: simulation agents do not have to be modified or accessed during the observation process~\citep{Kaminka:2002}.  However, Wilkins \textit{et al.} underline that the applicability of this method is limited: the observed property might not be inferred~\citep{Wilkins:2003}; it is often true in complex, \textit{i.e.}, in most of the real world, cases. Thus,  direct observation often remains the only available option.

The brute-force direct observation method is, for this particular and very simple toy-problem, linear in the number of simulation agents\footnote{Authors would like to thank anonymous reviewers for raising this issue.}, \textit{i.e.}, $\mathcal{C}(obs) \propto N$, while the model is exponential, \textit{i.e.}, $\mathcal{C}(model) \propto \alpha^N, \alpha > 1$. However, complex simulations generally involve non-linear observation problems~\citep{Veremme:2010}. Thus, improving direct observation appears to be a good lead to improve the efficiency of large-scale complex MABS.

In this paper, two non-brute-force direct observation methods, based on self-observation and statistical survey theory are introduced. An empirical comparison of the computational cost of these methods is performed and discussed on the presented case study.

\section{Filtrated direct observation of MABS}

\begin{figure}[ht]
	\begin{center}
\tikzstyle{blocvide}= [minimum width=0.2cm, minimum height=0.2cm,text centered]
\tikzstyle{blocdebase}= [draw, text centered,minimum height=1cm, rounded corners]

\tikzstyle{simulation}= [draw, dashed, text centered,minimum height=2.4cm, minimum width=8cm, rounded corners]

\tikzstyle{obs} = [blocdebase]

\pgfdeclarelayer{background}
\pgfdeclarelayer{foreground}
\pgfsetlayers{background,main,foreground}

\begin{tikzpicture}[->,>=stealth',shorten >=1pt,auto,semithick]
    
    \node[obs] (observateur) at (2cm, 0cm)  {$obs'$};
    
    \node[blocvide] (observation) at (5cm, 0cm)  {observation};
    
    \path (observateur) edge (observation) {};

    \node[blocvide] at (6cm, -2.7cm) {$\mathcal{A}$};
    \node[blocvide] at (2cm, -2.7cm) {$\mathcal{A}'$};
    
    \foreach \x in {1, 2,  3} {
    	\node [circle,draw]  (a) at (\x cm, -2cm){};
	\path (a) edge [<-]  node {} (observateur);
    }
    
     \foreach \x in {4, 5, 6, 7} {
    	\node [circle,draw]  (a) at (\x cm, -2cm){};
    }
		
	\begin{pgfonlayer}{background}
		\node[simulation] (A) at (4cm, -2cm){};
		\node[simulation, minimum height=2cm, minimum width=3cm] (Ap) at (2cm, -2cm){};
   	 \end{pgfonlayer}

	 	\draw [->,out=180] (A.west)+(0cm,-0.5cm) .. controls  +(-2cm,-0.5cm) and  +(-2cm,0.2cm) ..  (Ap.west);

	\node[blocvide] (filter) at (-0.6cm,-1.7cm) {$filter$};
    
\end{tikzpicture}
		\caption{Filtrated direct observation}
		\label{filtering-scheme}
		\end{center}
\end{figure}
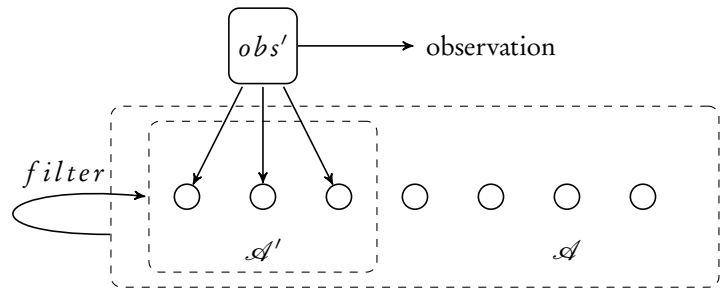

Basically, there are two ways to compute a direct observation: 
\begin{enumerate}
\item  a set of agents $\mathcal{A}$  (generally all the simulation agents that share the properties that have to be observed), statically defined, is probed by an observer agent, 
\item a subset $\mathcal{A}'$ of $\mathcal{A}$, computed at runtime, is probed (fig.~\ref{filtering-scheme}). 
\end{enumerate}

Formally, we consider an observation function
\begin{equation}
		obs : 2^\mathcal{A} \rightarrow \mathcal{I},
\end{equation}
where $\mathcal{A}$ is a set of agents and $\mathcal{I}$ represents the values that can be observed. We define $\mathcal{A}' \subseteq \mathcal{A}$ as the minimal subset of agents, dynamically defined by a set of constraints (\textit{e.g.}, in the case study, a unique constraint related to the position of the agent),  needed to compute $obs$ correctly. In other words,  $\mathcal{A}'$ is a set of agents such as
\begin{equation}
	\begin{array}{c}
  		obs'(\mathcal{A}') = obs(\mathcal{A}) \text{~and~}\\ 	 \nexists~\mathcal{A}''  \subset \mathcal{A}'~|~obs'(\mathcal{A}'') = obs(\mathcal{A}),
	\end{array}
\label{equ-filtre}
\end{equation}
where $obs'$ is a "simplified" observation function, \textit{i.e.}, that can be computed faster than  $obs$ because it is specific to $\mathcal{A}'$: 
\begin{equation}
	\begin{array}{c}
		\forall~\mathcal{A}'' \neq \mathcal{A}' \subseteq \mathcal{A},~obs'(\mathcal{A}') = obs(\mathcal{A}').
	\end{array}
\end{equation}

Thus, in the case study, the observation  function $obs(\mathcal{A})$ observes the position of each agent in $\mathcal{A}$ to count only the ones that are situated in $\mathcal{Z}$, while  an observation function $obs'(\mathcal{A}')$ would only returns $|\mathcal{A}'|$ because the agents of $\mathcal{A}'$ are by definition situated in $\mathcal{Z}$. 

To use $obs'$,  it is necessary to consider a filtering function $filter$ able to identify the subset $\mathcal{A}'$ (in the case study, this subset only contains the agents situated in $\mathcal{Z}$):
	 \begin{equation}
	 	\begin{array}{c}
			filter :  2^\mathcal{A} \rightarrow 2^\mathcal{A}~|~\forall \mathcal{A}', \mathcal{A}'' \subseteq \mathcal{A},\\
			\text{if~} filter(\mathcal{A}') = \mathcal{A}'', \text{~then~} \mathcal{A}'' \subseteq \mathcal{A}'.
			\end{array}
	\end{equation}
					
The goal is to define and implement a filtering function, such as the cost of the observation computation is reduced, \textit{i.e.},
	\begin{equation}
		\mathcal{C}(obs'(filter(\mathcal{A}))) < \mathcal{C}(obs(\mathcal{A})).
		\label{eq_complexite}
	\end{equation}

In the following section, two different implementations of this idea are presented.  

\section{Implementation of filtrated direct observation methods}

\subsection{Self-observation}

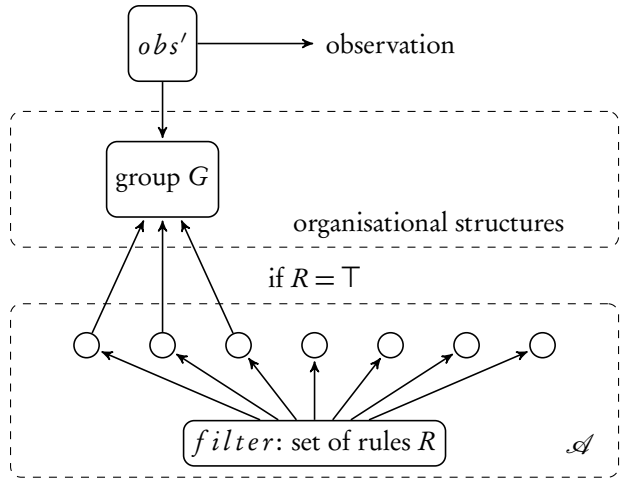
\begin{figure}[ht]
	\begin{center}
\tikzstyle{blocvide}= [minimum width=0.2cm, minimum height=0.2cm,text centered]
\tikzstyle{blocdebase}= [draw, text centered,minimum height=0.5cm, rounded corners]

\tikzstyle{simulation}= [draw, dashed, text centered,minimum height=2.3cm, minimum width=8cm, rounded corners]

\tikzstyle{obs} = [blocdebase]

\pgfdeclarelayer{background}
\pgfdeclarelayer{foreground}
\pgfsetlayers{background,main,foreground}

\begin{tikzpicture}[->,>=stealth',shorten >=1pt,auto,semithick]
    
    \node[obs,minimum height=1cm] (observateur) at (2cm, 2.6cm)  {$obs'$};
    
     \node[blocdebase] (regles) at (4cm, -2.7cm)  {$filter$: set of rules $R$};
     
     \node[blocdebase,minimum height=1cm] (groupe) at (2cm, 0.8cm)  {group $G$};
    
    \node[blocvide]  at (5.5cm, 0.2cm)  {organisational structures};
    
    \node[blocvide] (observation) at (5cm, 2.6cm)  {observation};
    
    \path (observateur) edge (observation) {};
    
     \path (observateur) edge (groupe) {};
     
      \node[blocvide] at (7.5cm, -2.7cm) {$\mathcal{A}$};
    
     \node[blocvide]  at (4cm, -0.5cm){if $R = \top$};
            
    \foreach \x in {1, 2,  3} {
    	\node [circle,draw]  (a) at (\x cm, -1.4cm){};
	\path (a) edge [->]  node {} (groupe);
	\path (regles) edge [->] node {} (a);
    }
    
     \foreach \x in { 4, 5 ,6, 7} {
    	\node [circle,draw]  (a) at (\x cm, -1.4cm){};
	\path (regles) edge  [->] node {} (a);
    }
		
	\begin{pgfonlayer}{background}
		\node[simulation] at (4cm, -2cm){};
		\node[simulation, minimum height=1.8cm] at (4cm, 0.8cm){};
   	 \end{pgfonlayer}

\end{tikzpicture}
		\caption{The self-observation based method}
		\label{selfObs-scheme}
		\end{center}
\end{figure}

The core idea of this method is to implement the filtering function in the simulation agents themselves. Then, using an organizational structure, denoted \textit{group}, allows to identify the set of agents that has to be observed. Thus, a group is defined as \textit{the set of agents that contains the sufficient and necessary information to compute an observation}. In other words, a group defines, for a given observation, the minimal set of agents that is mandatory to compute it. Agents observe themselves to determine if they have to join, leave or stay in a group. A filtering function is defined by a set of rules $R$,  that specifies the conditions under which an agent has to be observed, evaluated at each simulation step~(fig.~\ref{selfObs-scheme}). 

\begin{figure}[htp]
			\begin{center}
				\input{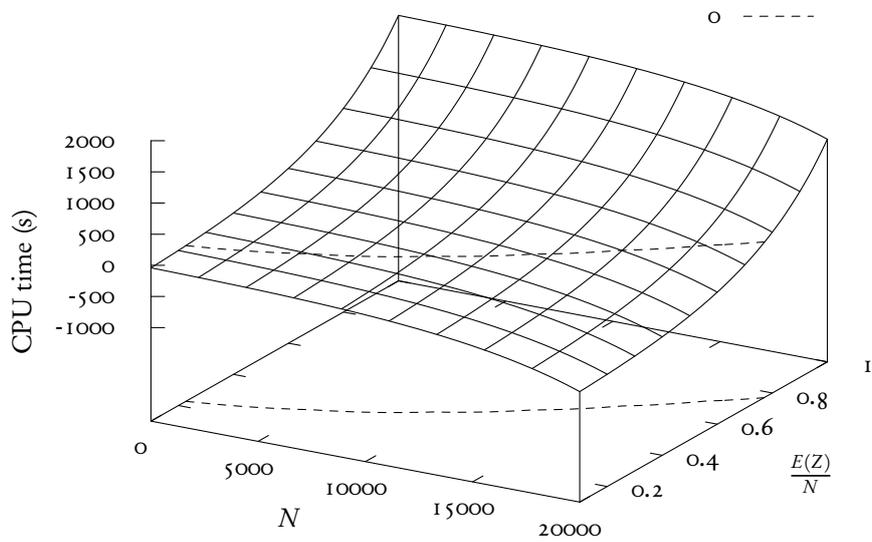}
				\caption{CPU time difference between simulations observed with self-observation  and  brute-force methods as a function of the number of agents in the simulation, $N$, and the mean rate of observed agents, $E(Z)/N$ (response surface estimation)}
				\label{image-naiveVSgroupes}
			\end{center}
		\end{figure}

Thus, in our case study, we consider a group $G$, that contains the agents situated in $\mathcal{Z}$. The following set of rules, defined here in natural language, is associated to each agent:  
		\begin{itemize}
			\item if the agent is in $\mathcal{Z}$ and does not belong to $G$, then the agent joins $G$,
			\item if the agent is not in $\mathcal{Z}$ and belongs to $G$, then the agent leaves $G$.
		\end{itemize}
		The observation system ($obs'$) only probes the agents of $G$. Figure~\ref{image-naiveVSgroupes} presents the CPU time difference between simulations observed with self-observation  and  brute-force methods as a function of the number of agents in the simulation, $N$, and the mean rate of observed agents, $E(Z)/N$. The dashed line represents the isoline $0$, \textit{i.e.}, the conditions for which there is no difference between the two methods. Thus, the area below this line maps the cases for which self-observation is faster.

\subsection{Statistical survey}
\label{section-survey}

	If the equation~\ref{equ-filtre} is rewritten as follows:
			\begin{equation}
				\begin{array}{c}
					obs(\mathcal{A}') \simeq obs(\mathcal{A}) \text{~and~}\\
					\nexists~\mathcal{A}'' \subset \mathcal{A}'~|~obs(\mathcal{A}'') \simeq obs(\mathcal{A}),
				\end{array}
			\end{equation}
			\textit{i.e.}, if imprecise observations are authorized, it becomes possible to filter the observed population on a statistical basis. As a result, we do not consider a specific observation function anymore as the set of agents returned by the filtering function is not necessary the \textit{set of agents that contains the sufficient and necessary information to compute the observation}. 
			
			 Statistical survey theory\footnote{Proofs of statistical survey theory results presented in this paper will not be given. Interested readers may refer to \citet{Bethlehem:2009} for an exhaustive presentation of sampling designs, estimator construction and variability estimation methods.} provides a formal ground to determine optimal sampling method and size of observed population sample. 
			 
			  Let once again consider our case study; we denote $n$ the size of the observed population, randomly sampled at each simulation step. An estimator of $Z$, denoted $\hat{Z}$ is constructed from this sample. Many estimator definitions can be found in the literature. In this case, as the population of simulation agents is homogeneous with respect to $E(Z)$ (all the agents have the same probability to be in $\mathcal{Z}$), $n$ is determined with the Horvitz-Thompson estimator~\citep{Horvitz:1952}~:			
			  \begin{equation}
				n^{-1} = \frac{d^2}{4S^2} + \frac{1}{N},
			\end{equation}	
			where $d$ is the maximal absolute error accepted for the observation and
			\begin{equation}
				S^2 \simeq (1 - \frac{E(Z)}{N}) \cdot \frac{E(Z)}{N}.
			\end{equation}
						
\begin{figure}[htp]
				\begin{center}
					\input{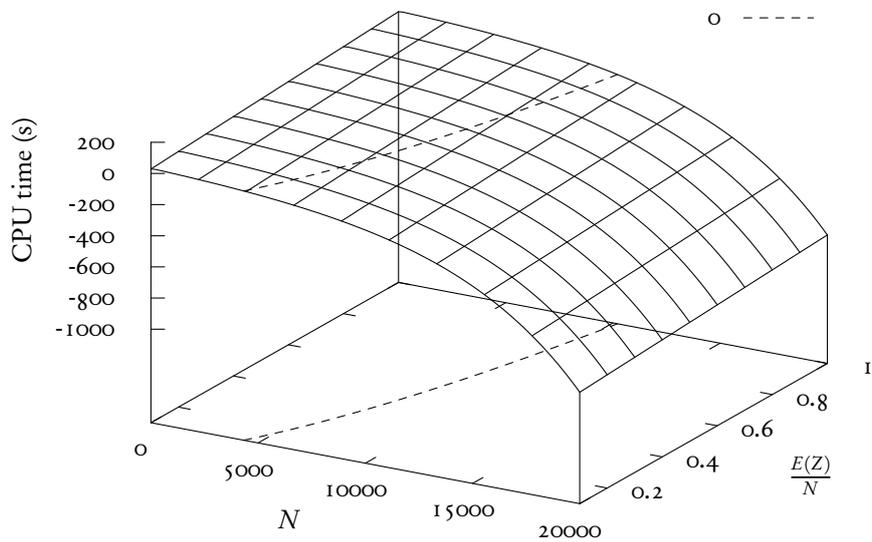}
					\caption{CPU time difference between simulations observed with statistical survey  ($d=0.08$) and  brute-force methods as a function of the number of agents in the simulation, $N$, and the mean rate of observed agents, $E(Z)/N$ (response surface estimation)}
					\label{image-naiveVSSample}
				\end{center}
			\end{figure}

			Impact on the computational cost is shown in figure~\ref{image-naiveVSSample}. The semantic is the same than figure~\ref{image-naiveVSgroupes}: the left area maps the cases for which the statistical survey based method is faster than brute-force method.
			
\subsection{Discussion}

\begin{figure}[htp]
	\begin{center}
		\begin{tikzpicture}[gnuplot]
\gpmonochromelines
\gpcolor{gp lt color border}
\node[gp node left] at (4.979,2.575) {A};
\node[gp node left] at (4.132,5.924) {B};
\node[gp node left] at (8.367,3.970) {C};
\gpcolor{gp lt color 0}
\gpsetlinetype{gp lt plot 0}
\gpsetlinewidth{1.00}
\draw[gp path] (2.015,3.020)--(2.486,3.085)--(2.956,3.150)--(3.427,3.245)--(3.898,3.340)%
  --(4.321,3.470)--(4.745,3.599)--(4.792,3.653)--(4.839,3.706)--(4.960,3.963)--(5.080,4.219)%
  --(5.226,4.529)--(5.373,4.839)--(5.519,5.149)--(5.665,5.459)--(5.722,5.581)--(5.780,5.703)%
  --(5.840,5.891)--(5.900,6.079)--(5.999,6.389)--(6.098,6.699)--(6.197,7.009)--(6.295,7.320);
\draw[gp path] (2.015,3.020)--(2.486,3.085)--(2.956,3.150)--(3.427,3.245)--(3.898,3.340)%
  --(4.321,3.470)--(4.745,3.599)--(4.792,3.653)--(4.839,3.706)--(4.960,3.963)--(5.080,4.219)%
  --(5.226,4.529)--(5.373,4.839)--(5.519,5.149)--(5.665,5.459)--(5.722,5.581)--(5.780,5.703)%
  --(5.840,5.891)--(5.900,6.079)--(5.999,6.389)--(6.098,6.699)--(6.197,7.009)--(6.295,7.320);
\draw[gp path] (8.091,1.738)--(7.979,2.049)--(7.868,2.359)--(7.765,2.479)--(7.661,2.600)%
  --(7.416,2.789)--(7.170,2.979)--(6.945,3.071)--(6.720,3.163)--(6.250,3.291)--(5.780,3.418)%
  --(5.309,3.503)--(4.839,3.587)--(4.827,3.593)--(4.815,3.599)--(4.827,3.617)--(4.839,3.636)%
  --(4.960,3.928)--(5.080,4.219)--(5.226,4.529)--(5.373,4.839)--(5.519,5.149)--(5.665,5.459)%
  --(5.722,5.581)--(5.780,5.703)--(5.840,5.891)--(5.900,6.079)--(5.999,6.389)--(6.098,6.699)%
  --(6.197,7.009)--(6.295,7.320);
\draw[gp path] (8.091,1.738)--(7.979,2.049)--(7.868,2.359)--(7.765,2.479)--(7.661,2.600)%
  --(7.416,2.789)--(7.170,2.979)--(6.945,3.071)--(6.720,3.163)--(6.250,3.291)--(5.780,3.418)%
  --(5.309,3.503)--(4.839,3.587)--(4.827,3.593)--(4.815,3.599)--(4.827,3.617)--(4.839,3.636)%
  --(4.960,3.928)--(5.080,4.219)--(5.226,4.529)--(5.373,4.839)--(5.519,5.149)--(5.665,5.459)%
  --(5.722,5.581)--(5.780,5.703)--(5.840,5.891)--(5.900,6.079)--(5.999,6.389)--(6.098,6.699)%
  --(6.197,7.009)--(6.295,7.320);
\gpcolor{gp lt color border}
\gpsetlinetype{gp lt border}
\draw[gp path] (2.014,7.320)--(2.014,1.737)--(10.485,1.737)--(10.485,7.320)--cycle;
\draw[gp path] (2.015,1.738)--(2.015,1.972);
\node[gp node center] at (2.015,1.338) { 0};
\draw[gp path] (2.015,7.320)--(2.015,7.086);
\draw[gp path] (4.133,1.738)--(4.133,1.972);
\node[gp node center] at (4.133,1.338) { 5000};
\draw[gp path] (4.133,7.320)--(4.133,7.086);
\draw[gp path] (6.250,1.738)--(6.250,1.972);
\node[gp node center] at (6.250,1.338) { 10000};
\draw[gp path] (6.250,7.320)--(6.250,7.086);
\draw[gp path] (8.367,1.738)--(8.367,1.972);
\node[gp node center] at (8.367,1.338) { 15000};
\draw[gp path] (8.367,7.320)--(8.367,7.086);
\draw[gp path] (10.485,1.738)--(10.485,1.972);
\node[gp node center] at (10.485,1.338) { 20000};
\draw[gp path] (10.485,7.320)--(10.485,7.086);
\node[gp node center] at (6.250,0.876) {$N$};
\draw[gp path] (2.015,1.738)--(2.249,1.738);
\node[gp node right] at (1.776,1.738) { 0};
\draw[gp path] (10.485,1.738)--(10.251,1.738);
\draw[gp path] (2.015,2.855)--(2.249,2.855);
\node[gp node right] at (1.776,2.855) { 0.2};
\draw[gp path] (10.485,2.855)--(10.251,2.855);
\draw[gp path] (2.015,3.971)--(2.249,3.971);
\node[gp node right] at (1.776,3.971) { 0.4};
\draw[gp path] (10.485,3.971)--(10.251,3.971);
\draw[gp path] (2.015,5.087)--(2.249,5.087);
\node[gp node right] at (1.776,5.087) { 0.6};
\draw[gp path] (10.485,5.087)--(10.251,5.087);
\draw[gp path] (2.015,6.203)--(2.249,6.203);
\node[gp node right] at (1.776,6.203) { 0.8};
\draw[gp path] (10.485,6.203)--(10.251,6.203);
\draw[gp path] (2.015,7.320)--(2.249,7.320);
\node[gp node right] at (1.776,7.320) { 1};
\draw[gp path] (10.485,7.320)--(10.251,7.320);
\node[gp node center] at (0.581,4.529) {$\frac{E(Z)}{N}$};
\gpdefrectangularnode{gp plot 1}{\pgfpoint{2.014cm}{1.737cm}}{\pgfpoint{10.485cm}{7.320cm}}
\draw[gp path] (2.014,7.320)--(2.014,1.737)--(10.485,1.737)--(10.485,7.320)--cycle;
\end{tikzpicture}
		\caption{Map of the fastest observation methods (response surface estimation); A: self-observation, B: brute-force, C: statistical survey ($d=0.008$)}
		\label{image-bilanObservation}
	\end{center}
\end{figure}
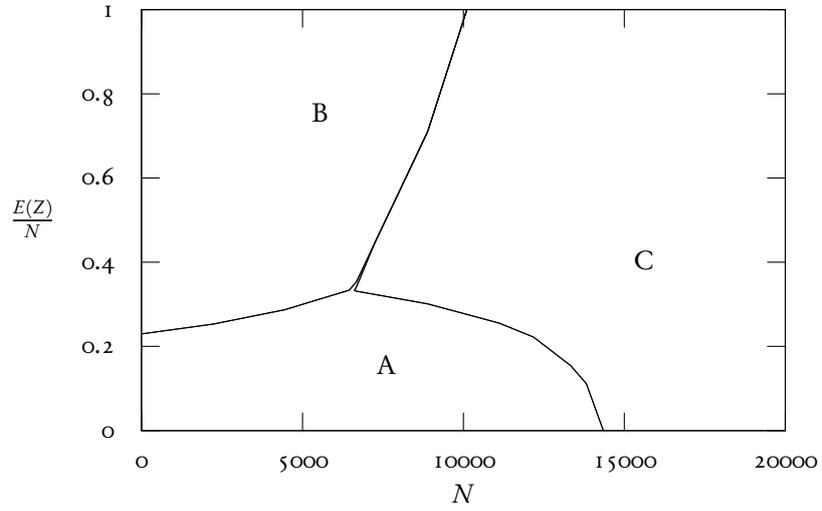

Figure~\ref{image-bilanObservation} sums up the previous results qualitatively: conditions for which it is preferable to use one method over another are identified. These results are specific to our case study and its implementation; however, they highlight that the choice of an observation method is not trivial and that the performance of the different available methods should be analyzed on a set of simulations before using the model in a production context.

\begin{figure}[htp]
	\begin{center}
		\begin{tikzpicture}[gnuplot]
\gpmonochromelines
\gpcolor{gp lt color border}
\gpsetlinetype{gp lt border}
\gpsetlinewidth{1.00}
\draw[gp path] (1.688,0.985)--(1.778,0.985);
\draw[gp path] (11.947,0.985)--(11.857,0.985);
\draw[gp path] (1.688,1.460)--(1.778,1.460);
\draw[gp path] (11.947,1.460)--(11.857,1.460);
\draw[gp path] (1.688,1.703)--(1.778,1.703);
\draw[gp path] (11.947,1.703)--(11.857,1.703);
\draw[gp path] (1.688,1.819)--(1.868,1.819);
\draw[gp path] (11.947,1.819)--(11.767,1.819);
\node[gp node right] at (1.504,1.819) { 1};
\draw[gp path] (1.688,2.178)--(1.778,2.178);
\draw[gp path] (11.947,2.178)--(11.857,2.178);
\draw[gp path] (1.688,2.652)--(1.778,2.652);
\draw[gp path] (11.947,2.652)--(11.857,2.652);
\draw[gp path] (1.688,2.896)--(1.778,2.896);
\draw[gp path] (11.947,2.896)--(11.857,2.896);
\draw[gp path] (1.688,3.011)--(1.868,3.011);
\draw[gp path] (11.947,3.011)--(11.767,3.011);
\node[gp node right] at (1.504,3.011) { 10};
\draw[gp path] (1.688,3.371)--(1.778,3.371);
\draw[gp path] (11.947,3.371)--(11.857,3.371);
\draw[gp path] (1.688,3.845)--(1.778,3.845);
\draw[gp path] (11.947,3.845)--(11.857,3.845);
\draw[gp path] (1.688,4.089)--(1.778,4.089);
\draw[gp path] (11.947,4.089)--(11.857,4.089);
\draw[gp path] (1.688,4.204)--(1.868,4.204);
\draw[gp path] (11.947,4.204)--(11.767,4.204);
\node[gp node right] at (1.504,4.204) { 100};
\draw[gp path] (1.688,4.563)--(1.778,4.563);
\draw[gp path] (11.947,4.563)--(11.857,4.563);
\draw[gp path] (1.688,5.038)--(1.778,5.038);
\draw[gp path] (11.947,5.038)--(11.857,5.038);
\draw[gp path] (1.688,5.281)--(1.778,5.281);
\draw[gp path] (11.947,5.281)--(11.857,5.281);
\draw[gp path] (1.688,5.397)--(1.868,5.397);
\draw[gp path] (11.947,5.397)--(11.767,5.397);
\node[gp node right] at (1.504,5.397) { 1000};
\draw[gp path] (1.688,5.756)--(1.778,5.756);
\draw[gp path] (11.947,5.756)--(11.857,5.756);
\draw[gp path] (1.688,0.985)--(1.688,1.165);
\draw[gp path] (1.688,5.756)--(1.688,5.576);
\node[gp node center] at (1.688,0.677) { 2000};
\draw[gp path] (2.828,0.985)--(2.828,1.165);
\draw[gp path] (2.828,5.756)--(2.828,5.576);
\node[gp node center] at (2.828,0.677) { 4000};
\draw[gp path] (3.968,0.985)--(3.968,1.165);
\draw[gp path] (3.968,5.756)--(3.968,5.576);
\node[gp node center] at (3.968,0.677) { 6000};
\draw[gp path] (5.108,0.985)--(5.108,1.165);
\draw[gp path] (5.108,5.756)--(5.108,5.576);
\node[gp node center] at (5.108,0.677) { 8000};
\draw[gp path] (6.248,0.985)--(6.248,1.165);
\draw[gp path] (6.248,5.756)--(6.248,5.576);
\node[gp node center] at (6.248,0.677) { 10000};
\draw[gp path] (7.387,0.985)--(7.387,1.165);
\draw[gp path] (7.387,5.756)--(7.387,5.576);
\node[gp node center] at (7.387,0.677) { 12000};
\draw[gp path] (8.527,0.985)--(8.527,1.165);
\draw[gp path] (8.527,5.756)--(8.527,5.576);
\node[gp node center] at (8.527,0.677) { 14000};
\draw[gp path] (9.667,0.985)--(9.667,1.165);
\draw[gp path] (9.667,5.756)--(9.667,5.576);
\node[gp node center] at (9.667,0.677) { 16000};
\draw[gp path] (10.807,0.985)--(10.807,1.165);
\draw[gp path] (10.807,5.756)--(10.807,5.576);
\node[gp node center] at (10.807,0.677) { 18000};
\draw[gp path] (11.947,0.985)--(11.947,1.165);
\draw[gp path] (11.947,5.756)--(11.947,5.576);
\node[gp node center] at (11.947,0.677) { 20000};
\draw[gp path] (1.688,5.756)--(1.688,0.985)--(11.947,0.985)--(11.947,5.756)--cycle;
\node[gp node center,rotate=-270] at (0.246,3.370) {CPU time (s)};
\node[gp node center] at (6.817,0.215) {Number of agents};
\node[gp node right] at (5.534,5.379) {no observation};
\gpcolor{gp lt color 0}
\gpsetlinetype{gp lt plot 0}
\draw[gp path] (5.718,5.379)--(6.634,5.379);
\draw[gp path] (1.688,1.344)--(2.258,1.798)--(2.828,1.741)--(3.398,1.970)--(3.968,1.955)%
  --(4.538,2.117)--(5.108,2.140)--(5.678,2.276)--(6.248,2.318)--(6.818,2.440)--(7.387,2.476)%
  --(7.957,2.581)--(8.527,2.623)--(9.097,2.709)--(9.667,2.760)--(10.237,2.830)--(10.807,2.889)%
  --(11.377,2.936)--(11.947,2.998);
\gpsetpointsize{4.00}
\gppoint{gp mark 1}{(1.688,1.344)}
\gppoint{gp mark 1}{(2.258,1.798)}
\gppoint{gp mark 1}{(2.828,1.741)}
\gppoint{gp mark 1}{(3.398,1.970)}
\gppoint{gp mark 1}{(3.968,1.955)}
\gppoint{gp mark 1}{(4.538,2.117)}
\gppoint{gp mark 1}{(5.108,2.140)}
\gppoint{gp mark 1}{(5.678,2.276)}
\gppoint{gp mark 1}{(6.248,2.318)}
\gppoint{gp mark 1}{(6.818,2.440)}
\gppoint{gp mark 1}{(7.387,2.476)}
\gppoint{gp mark 1}{(7.957,2.581)}
\gppoint{gp mark 1}{(8.527,2.623)}
\gppoint{gp mark 1}{(9.097,2.709)}
\gppoint{gp mark 1}{(9.667,2.760)}
\gppoint{gp mark 1}{(10.237,2.830)}
\gppoint{gp mark 1}{(10.807,2.889)}
\gppoint{gp mark 1}{(11.377,2.936)}
\gppoint{gp mark 1}{(11.947,2.998)}
\gppoint{gp mark 1}{(6.176,5.379)}
\gpcolor{gp lt color border}
\node[gp node right] at (5.534,5.071) {brute-force observation};
\gpcolor{gp lt color 1}
\gpsetlinetype{gp lt plot 1}
\draw[gp path] (5.718,5.071)--(6.634,5.071);
\draw[gp path] (1.688,0.985)--(1.688,2.337)--(2.828,2.907)--(3.968,3.446)--(5.108,3.868)%
  --(6.248,4.201)--(7.387,4.486)--(8.527,4.756)--(9.667,4.962)--(10.807,5.165)--(11.947,5.332);
\gppoint{gp mark 2}{(1.688,2.337)}
\gppoint{gp mark 2}{(2.828,2.907)}
\gppoint{gp mark 2}{(3.968,3.446)}
\gppoint{gp mark 2}{(5.108,3.868)}
\gppoint{gp mark 2}{(6.248,4.201)}
\gppoint{gp mark 2}{(7.387,4.486)}
\gppoint{gp mark 2}{(8.527,4.756)}
\gppoint{gp mark 2}{(9.667,4.962)}
\gppoint{gp mark 2}{(10.807,5.165)}
\gppoint{gp mark 2}{(11.947,5.332)}
\gppoint{gp mark 2}{(6.176,5.071)}
\gpcolor{gp lt color border}
\node[gp node right] at (5.534,4.763) {filtrated observation};
\gpcolor{gp lt color 2}
\gpsetlinetype{gp lt plot 2}
\draw[gp path] (5.718,4.763)--(6.634,4.763);
\draw[gp path] (1.688,1.649)--(2.258,2.069)--(2.828,2.157)--(3.398,2.443)--(3.968,2.600)%
  --(4.538,2.809)--(5.108,2.939)--(5.678,3.114)--(6.248,3.244)--(6.818,3.398)--(7.387,3.512)%
  --(7.957,3.614)--(8.527,3.738)--(9.097,3.849)--(9.667,3.953)--(10.237,3.973)--(10.807,3.983)%
  --(11.377,3.993)--(11.947,4.004);
\gppoint{gp mark 3}{(1.688,1.649)}
\gppoint{gp mark 3}{(2.258,2.069)}
\gppoint{gp mark 3}{(2.828,2.157)}
\gppoint{gp mark 3}{(3.398,2.443)}
\gppoint{gp mark 3}{(3.968,2.600)}
\gppoint{gp mark 3}{(4.538,2.809)}
\gppoint{gp mark 3}{(5.108,2.939)}
\gppoint{gp mark 3}{(5.678,3.114)}
\gppoint{gp mark 3}{(6.248,3.244)}
\gppoint{gp mark 3}{(6.818,3.398)}
\gppoint{gp mark 3}{(7.387,3.512)}
\gppoint{gp mark 3}{(7.957,3.614)}
\gppoint{gp mark 3}{(8.527,3.738)}
\gppoint{gp mark 3}{(9.097,3.849)}
\gppoint{gp mark 3}{(9.667,3.953)}
\gppoint{gp mark 3}{(10.237,3.973)}
\gppoint{gp mark 3}{(10.807,3.983)}
\gppoint{gp mark 3}{(11.377,3.993)}
\gppoint{gp mark 3}{(11.947,4.004)}
\gppoint{gp mark 3}{(6.176,4.763)}
\gpcolor{gp lt color border}
\gpsetlinetype{gp lt border}
\draw[gp path] (1.688,5.756)--(1.688,0.985)--(11.947,0.985)--(11.947,5.756)--cycle;
\gpdefrectangularnode{gp plot 1}{\pgfpoint{1.688cm}{0.985cm}}{\pgfpoint{11.947cm}{5.756cm}}
\end{tikzpicture}
		\caption{CPU times (log scale) needed to compute observed and unobserved simulations for $E(Z)=N/5$, filtrated observation method being dynamically adapted to the context}
		\label{image-bestObs}
	\end{center}
\end{figure}
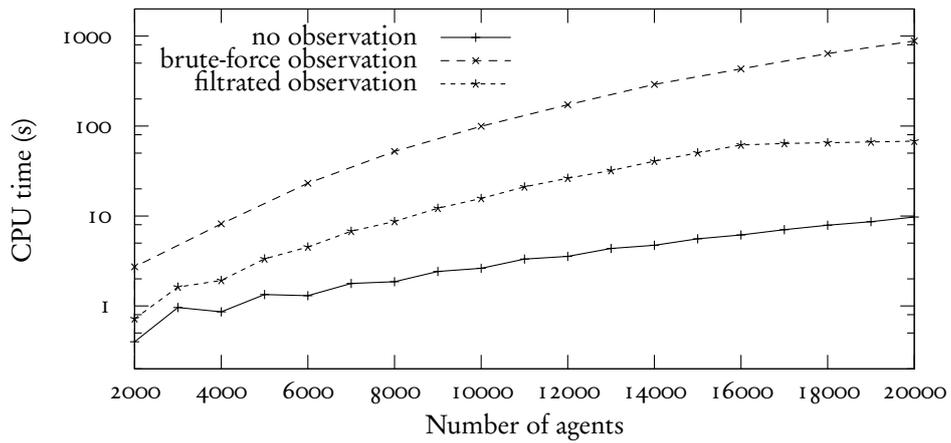

In a given context, knowing the map of the fastest observation methods allows to dynamically adapt the observation method to use the most efficient one. Impact of dynamic adaptation of the observation method  on CPU time is presented in the context of the first example (cf. fig.~\ref{fig1}) in figure~\ref{image-bestObs}. 

 
\section{Conclusion and perspectives}

Observation methods presented in this paper allow, under specific conditions identified on a simple case study, to reduce significantly the computational cost of MABS composed of numerous agents.  However, filtering is not the only option. Considering that imprecise observations are acceptable, while a precision level is guaranteed, the optimal observation frequency could be determined from  the observed property variation. Roughly, the more the variation, the more the observation frequency. However, MABS are used to simulate complex systems with nonlinear dynamics. Dynamic adaptation of observation frequency could be an interesting lead to reduce MABS computational cost.

Moreover, in the statistical survey  based method (cf. section~\ref{section-survey}), we consider a simple random sampling method, assuming the population is homogeneous. Real world MABS often involve heterogeneous agents for which the distribution of observed individual properties is not uniform. Clever sampling methods, \textit{e.g.}, a stratified random sampling approach, should then be used. In very complex cases, a machine learning system should be implemented to analyze the impact of sampling method properties on the observation quality and computational cost, and determine the optimal ones. Similarly, the organizational model used to implement the self-observation based method is very simple: the only organizational structure that is defined is the "group". Using a more comprehensive one, \textit{e.g.}, AGR~\citep{Ferber:1998}, would allow to consider very complex and fine observations. 

From a methodological point of view, authors experimented that setting up an observation method, generally improves the design of simulation validity metrics. Indeed, it forces simulation designers and users to explicitly define local and global observed properties and their sufficient and necessary conditions of observability,  and then the validity constraints over them.

While this paper focuses on reducing the complexity of observation, many published works concentrated on agent interactions by dynamically scaling up and down simulated entities or using more structured interaction artifacts~\citep{Gaud:2008,Razavi:2011,Parunak:2012}. Together, these approaches should lead to the conception of highly efficient large-scale MABS simulators.

\bibliographystyle{apalike}
\bibliography{../../Biblio}

\end{document}